\documentclass[11pt,a4paper]{article}
\usepackage[utf8]{inputenc}
\usepackage{amssymb, amsbsy} % simbolos
\usepackage{amsfonts}
\usepackage[centertags]{amsmath}
\usepackage{mathtools}
\mathtoolsset{showonlyrefs=true} % solo se muestra la etiqueta de la eq si se cita
\usepackage{amsthm}
\usepackage{mathabx} %para el sqbullet
\usepackage{nccmath}
\usepackage{array}
\usepackage{graphicx} % figuras
\usepackage{subfigure}  % subfiguras
\usepackage[left=2cm,right=2cm,top=2cm,bottom=2cm]{geometry}
\usepackage{upgreek} % para poner letras griegas sin cursiva
\usepackage{cancel} % para tachar
\usepackage{mathdots} % para el comando \iddots
\usepackage{mathrsfs} % para formato de letra
\usepackage{stackrel} % para el comando \stackbin
\usepackage{comment} % para comentar varias lineas a la vez
\usepackage{bm} % para negrita en formulas
\usepackage{newlfont}
\usepackage{fancyhdr}
\usepackage{float} %para colocar las imagenes con H donde mandes
\usepackage{subfigure}% subcaptions for subfigures
\usepackage{longtable}% for tables in more than one page
\usepackage{booktabs,makecell} %for multiple lines in the header of the tables
\usepackage[dvipsnames]{xcolor}
\usepackage[final,pdftex,bookmarks=true,bookmarksopen=true,breaklinks=true,colorlinks=true,bookmarksnumbered=true,pdfsubject={ },pdfauthor={Ar\'is Fanjul Hevia}]{hyperref} 

\hypersetup{urlcolor=blue,linkcolor=blue,citecolor=blue}
\usepackage{natbib}
\usepackage{setspace}
\usepackage[toc,page]{appendix}
\usepackage{longtable}
\usepackage{multirow}
\usepackage{caption}
\usepackage{enumerate}
\usepackage{soul} % en vez de soul

\usepackage{authblk} % permite poner afiliacion
\title{\textbf{A new test for assessing the covariate effect\\ in ROC curves}}

\author[1]{Ar\'is Fanjul-Hevia}
\author[2]{Juan Carlos Pardo-Fern\'andez}
\author[3]{Wenceslao Gonz\'alez-Manteiga}
\affil[1]{Departamento de Estad\'istica e Investigación Operativa y Didáctica de la Matemática, Universidad de Oviedo}
\affil[2]{Departamento de Estat\'istica e Investigaci\'on Operativa  and CITMAga, Universidade de Vigo}
\affil[3]{Departamento de Estat\'istica, An\'alise Matem\'atica e Optimizaci\'on and CITMAga, Universidade de Santiago de Compostela}
\date{}
\definecolor{
}{rgb}{0.,0.5,0.}
\definecolor{darkred}{RGB}{138,0,0}

\graphicspath{ {Figuras/} }

\begin{document}

\maketitle

%=================================================================
\begin{abstract}
The ROC curve is a statistical tool that analyses the accuracy of a diagnostic test in which a variable is used to decide whether an individual is healthy or not. 
Along with that diagnostic variable it is usual to have information of some other covariates. In some situations it is advisable to incorporate that information into the study, as the performance of the ROC curves can be affected by them. Using the covariate-adjusted, the covariate-specific or the pooled ROC curves we discuss  how to decide if we can exclude the covariates from our study or not, and the implications this may have in further analyses of the ROC curve. A new test for comparing the covariate-adjusted and the pooled ROC curve is proposed, and the problem is illustrated by analysing a real database.
\end{abstract}


%=================================================================
\section{Introduction}
\label{sec:1}
%=================================================================

The ROC (Receiver Operating Characteristic) curve is a statistical tool that analyses the accuracy of a classification test. It has special relevance in medical studies where the aim is to differentiate the healthy from the diseased patients.  It combines the notion of sensitivity and specificity (the probabilities of correctly diagnosing a diseased and a healthy patient, respectively) of certain diagnostic variables in order to determine how well they are discriminating both populations. For more information on the construction of ROC curves we refer to the books of \cite{Pepe2003}, \cite{Krzanowski2009} and
\cite{Nakas2023}.

In a practical situation apart from those diagnostic markers it is usual to have some other covariates (such as the {blood pressure}, the {age} or the {body mass index} of the patients). Three different situations can arise when incorporating  covariates to the ROC curve analysis: in the first one the performance of the ROC curve changes with the value of the covariates (and with it, its discriminatory capability); in the second, the covariates affect the distribution of the diagnostic markers, but not their discriminatory capability; in the last one, the covariates do not affect the ROC curve in any way.  For some examples of these situations, check \cite{Pardo-Fernandez2014}. In this paper we propose a test for distinguishing if we are in the second or in the third type of situation.

In the literature there are three different curves that we can use in this context, three different ways of modelling this covariate information. The first one is the already mentioned ROC curve, that we will be calling the \textit{pooled ROC curve} because it uses all the pooled data disregarding any covariate information. Let $Y^F$ and $Y^G$ be the continuous  diagnostic variables in the diseased and the healthy populations, respectively. Let $X^F$ and $X^G$ be covariates associated to those populations and $R_X$ the intersection of the supports of $X^F$ and $X^G$ (assumed to be non-empty). Then the pooled ROC curve can be defined as
\begin{eqnarray}
ROC(p) &=& 1- F(G^{-1}(1-p)), \;  p\in (0,1),
 \label{eq:ROCdef}
\end{eqnarray}
where $F(y) = P(Y^F \leq y)$ and $G(y) = P(Y^G \leq y)$ are the cumulative distribution functions of $Y^F$ and $Y^G$, respectively, and $G^{-1}$ is the quantile function associated to $G$.

The second option is the \textit{conditional} or the \textit{covariate-specific ROC curve}. Given a fixed value of the covariate, $x \in R_X$, it is defined as 
\begin{eqnarray}
ROC^x(p) = 1- F(G^{-1} (1-p \mid x)\mid x), \; p \in (0,1),
\label{eq:ROCxdef}
\end{eqnarray}
where $F(\cdot\mid x)$  and  $G(\cdot\mid x)$ are the cumulative distribution functions of the diagnostic variable in the diseased and healthy population, respectively, conditioned to the value $x$. It may lead to different curves depending on $x$ and therefore it is the most convenient curve to use when the performance of the diagnostic variables changes depending on the value of the covariate.

The last alternative is the \textit{covariate-adjusted ROC curve} (\textit{AROC} curve), which was first introduced by \cite{JanesPepe2009} and was further used in  \cite{Janes2009}, \cite{Rodriguez-Alvarez2011} and \cite{InacioRdzAlvz2022}. It is defined as the ROC curve  that is obtained when the thresholds used for the classification are covariate-specific:%, and it can be viewed as a weighted average of conditional ROC curves:

\begin{eqnarray}
AROC(p) = P(Y^F>  G^{-1}(1-p\mid X^F)), \; p \in (0,1).
\label{eq:AROCdef}
\end{eqnarray}

The AROC curve can also be viewed as a weighted average of conditional ROC curves as
%(the weight depending on the distribution of $X^F$):
\begin{eqnarray}
AROC(p) = \int ROC^x(p) dF^{X}(x), \; p \in (0,1),
\label{eq:AROCdef2}
\end{eqnarray}
where $F^{X} (x) = P(X^F \leq x)$ is the cumulative distribution function of the covariate in the diseased population.

This curve could be particularly useful for giving a summary of the performance of the conditional ROC curve when sample sizes are not large enough for the conditional ROC curve to be estimated accurately. Moreover,  when the covariate affects the diagnostic variables but not their discriminatory accuracy (i.e., when the performance of the diagnostic marker is the same across populations with different values of the covariate), the AROC curve coincides with the conditional ROC curve. 

A first step towards deciding which of these three kinds of curves should be used for a particular situation would be to test whether the conditional ROC curves do not change along with the values of the covariates. 
%, meaning $ROC^z(p)= ROC^x(p) \; \forall x\in R_X$, with $ p\in (0,1)$, for a certain fixed value $z\in R_X$. 
Given that these curves would coincide with the AROC curve, this is equivalent to testing if $ROC^x(p)=AROC(p), \; p \in (0,1),$  for all $x\in R_X$. %\cite{Rodriguez-Alvarez2011} proposed a test for dealing with the this test in a case with a unidimensional covariate.
The literature regarding this problem is scarce. \cite{Rodriguez-Alvarez2011} gave a proposal for this test for the case with a unidimensional covariate. In a related context, \cite{Rodriguez-Alvarez2018}  proposed an inferential procedure for testing the effect of covariates over the conditional ROC curve employing generalized additive models.

However, even if that hypothesis of equality among the conditional and the adjusted ROC curves holds it does not mean that we can disregard the use of the covariates, because they can still have an effect on the diagnostic variables (which can lead, for example, to different optimal cut-off points). Thus, it is  of practical  interest to test for the null hypothesis
\begin{eqnarray}
H_0: AROC(p)=ROC(p), \; p \in (0,1),
\label{eq:TestH02}
\end{eqnarray}
versus the general alternative $H_1: H_0$ is not true. If this hypothesis were rejected, the AROC curve should be employed. If not, the pooled ROC curve can be used (provided that it has already been proven that the covariate has no effect on the discriminating capability of the diagnostic variables). The main objective of this paper is to propose a test for this problem. To the best of our knowledge, at this point there are no alternative methods in the literature that tackle this particular issue.

This piece of research is organised as follows. 
In Section~\ref{sec:2} of this paper a new non-parametric methodology  for testing the equality of the covariate-adjusted ROC curve and the pooled ROC curve is introduced. It includes a bootstrap algorithm to approximate the distribution of the statistic proposed. In Section~\ref{sec:3} %simulations are carried out for four different scenarios that capture the different relationships that the pooled ROC curve, the conditional ROC curve and the covariate-adjusted ROC curve may have. 
the different relationships between the pooled ROC curve, the conditional ROC curve and the covariate-adjusted ROC curve are further illustrated via different scenarios, scenarios then used to carry out a simulation study. 
This is followed by a real-data application for illustration purposes in Section~\ref{sec:4}. The last section is left for the discussion.
 The code for implementing this methodology \citep[written in the sowftware][]{ManR} is provided in Appendix~\ref{sec:Ap}.

%=================================================================
\section{Methodology}
\label{sec:2}
%=================================================================

The objective in this section is to propose a nonparametric test to decide whether the covariate at hand has an effect on the performance of the diagnostic variable or not. In other words, the aim is to test  whether the AROC curve coincides with the pooled ROC curve, as established in \eqref{eq:TestH02}.

Before presenting the test statistic that we will be using to tackle this problem, let us introduce the estimators of both curves in the case of having a unidimensional covariate. Note that there are several alternatives in the literature for these estimators, and that other approaches could be adapted to this problem. Check \cite{gonccalves2014} for a review of the existing methodologies for the estimation of the pooled ROC curve, and \cite{Pardo-Fernandez2014} for a review of the estimation procedures of the conditional and the AROC curves.

Let $\{(X_{i}^F,Y_{i}^F)\}_{i=1}^{n^F}$ be an i.i.d. sample from the distribution of $(X^F,Y^F)$  and $\{(X_{i}^G,Y_{i}^G)\}_{i=1}^{n^G}$ an i.i.d. sample from the distribution of $(X^G,Y^G)$. We will assume that $(X^F,Y^F)$ and $(X^G,Y^G)$ are independent.
For the estimation of the pooled ROC curve we will be considering the empirical estimator \citep{Hsieh1996}, which consists on plugging the empirical estimates of $F$ and $G^{-1}$ in \eqref{eq:ROCdef}, obtaining
\begin{eqnarray}
\widehat{ROC}(p) = 1- \hat{F}(\hat{G}^{-1}(1-p)), \; p\in (0,1),
\label{eq:ROCestemp}
\end{eqnarray}
where $\hat{F}(t) = (n^F)^{-1} \sum_{i=1}^{n^F} I(Y_{i}^F\leq t)$, $\hat{G}(t) = (n^G)^{-1} \sum_{i=1}^{n^G} I(Y_{i}^G\leq t)$ are the empirical distribution functions and $\hat{G}^{-1}(p) = \inf\{ t: \hat{G}(t) \geq p\}$ is the empirical quantile distribution function.

For the estimation of the AROC curve, following the ideas in \cite{Rodriguez-Alvarez2011}, we use nonparametric location-scale regression models to express the relationship between the diagnostic marker and the covariate. More specifically, for $D\in \{F,G\}$, let
\begin{eqnarray}
\label{eq:locscale}
Y^D = \mu^F(X^D) + \sigma^D(X^D)\varepsilon^D, 
\end{eqnarray}
where $X^D$ is the covariate associated with $Y^D$, $\mu^D(\cdot) = \mathbb{E}(Y^D\mid X^D=\cdot)$ and $(\sigma^D)^2(\cdot) = \mathbb{V}\text{ar}(Y^D \mid  X^D=\cdot)$ are the conditional mean and the conditional variance functions (both of them unknown smooth functions), and the error $\varepsilon^D$ is independent of $X^D$ and has cumulative distribution function $H^D$.  This approach was also used in \cite{Gonzalez-Manteiga2011a} to estimate the conditional ROC curve defined in \eqref{eq:ROCxdef} in terms of those error cumulative distribution functions, $H^F$ and $H^G$. 
The advantage of using location-scale regression models is that the conditional quantile can be expressed in terms of the error distribution, since $G^{-1} (p\mid x) = \mu^G(x)+\sigma^G(x)(H^G)^{-1}(p)$, with $p\in (0,1)$. This  allows us to expressed the AROC curve definied in \eqref{eq:AROCdef} as
\begin{eqnarray}
AROC(p) &=& P( Y^F > \mu^G(X^F)+\sigma^G(X^F)(H^G)^{-1}(1-p))\\
 &=& P\left(  H^G \left( \frac{Y^F - \mu^G(X^F)}{\sigma^G(X^F)}\right) > 1- p\right).
\end{eqnarray}

From the previous expression yields the following estimator of the AROC curve:
\begin{eqnarray}
%\widehat{AROC}(p)&=& \frac{1}{n^F} \sum_{i=1}^{n^F} I\left( \frac{Y_i^F - \hat{\mu}^G(X_i^F)}{\hat{\sigma}^G(X_i^F)} > (\hat{H}^G)^{-1}(1-p)\right), \; p \in (0,1),
\widehat{AROC}(p)&=& \frac{1}{n^F} \sum_{i=1}^{n^F} I\left( \hat{H}^G\left(\frac{Y_i^F - \hat{\mu}^G(X_i^F)}{\sigma^G(X_i^F)}\right) > 1-p\right), \; p \in (0,1),
\label{eq:AROCestim}
\end{eqnarray}
where, for $D\in \{F, G\},$
\begin{itemize}
\item  $\hat{H}^D(y) = (n^D)^{-1} \sum_{i=1}^{n^D} I(\hat{\varepsilon}_i^{D} \leq y)$, 
\item $\hat{\varepsilon}_i^{D} = \frac{Y_i^{D} -\hat{\mu}^D(X_i^{D})}{\hat{\sigma}^D (X_i^{D})}$ for $i\in\{1,\cdots,n^D\}$,
\item $\hat{\mu}^D(x) = \sum_{i=1}^{n_D}  W_i^{D} (x, g^D) Y_i^{D}$  is a nonparametric estimator of $\mu^D(x)$  based on local weights $W_i^{D}(x,g^D)$  depending on a bandwidth parameter $g^D$,
\item {\small{$(\hat{\sigma}^D)^2(x) = \sum_{i=1}^{n^D}  W_i^{D} (x, g^D) [Y_i^{D}-\hat{\mu}^D(X_i^{D})]^2$}}  is a nonparametric estimator of  {\small{$(\sigma^D)^2(x)$}}. For simplicity we take the same bandwidth parameter $g^D$ that is used for the estimation of the regression function $\hat{\mu}^D(x)$,
\item  $W_i^{D} (x,g^D) = \frac{\kappa_{g^D}(x-X_i^D)}{\sum_{l=1}^{n^D} \kappa_{g^D}(x-X_l^D)}$,  for $i\in\{1,\cdots,n^D\}$,  are Nadaraya-Watson-type weights, where $\kappa_{g^D}(\cdot)=$ $\kappa(\cdot/g^D)/g^D$ and $\kappa$ is the kernel (typically, a probability density function).
\end{itemize}

%\[H_0^2: AROC(p) = ROC(p) , \; \text{ for all } \; p\in (0,1), \]
%versus
%\[H_1^2: AROC(p) \neq ROC(p), \; \text{ for some } p \in (0,1). \]

Now we are in a position to handle the test presented in \eqref{eq:TestH02}. 
For that purpose we will compare the estimators of the pooled and the covariate-adjusted ROC curves through a test statistic of the form
\begin{eqnarray}
S_{\psi} = \psi \left( \widehat{ROC}(p) - \widehat{AROC}(p) \right),
\label{eq:Sphi}
\end{eqnarray}
where $\psi$ is a continuous function chosen to measure the distance between those curves. 
In particular we take three different distance functions, $\psi_{L_1}$, $\psi_{L_2}$ and $\psi_{KS}$, two based on the $L_1$ and the $L_2$ measures and the other on the Kolmogorov-Smirnov criterion based on the supreme. Similar distance functions have been used to measure the difference between ROC curves in \cite{Martinez-Camblor2011b, Martinez-Camblor2013a, Fanjul-Hevia2019}. This leaves us with
\begin{itemize}
\item $S_{\psi_{L_1}} = \int | \widehat{ROC}(p)  - \widehat{AROC}(p) | dp$,
\item $S_{\psi_{L_2}} = \int ( \widehat{ROC}(p)  - \widehat{AROC}(p) )^2 dp$,
\item $S_{\psi_{KS}} = \sup_{p \in (0,1)} \left| \widehat{ROC}(p)  - \widehat{AROC}(p) \right|$.
\end{itemize}
These test statistics will take values close to zero when under the null hypothesis, and large positive values  when under the alternative hypothesis.

%To avoid the dependency problems that arise from estimating both curves using the same data, the original sample, conformed by $\{(X_i^F,Y_i^F)\}_{i=1}^{n^F}$ and $\{(X_i^G,Y_i^G)\}_{i=1}^{n^G}$, was divided (randomly and evenly) in two sets. One of those, $\{Y_{R,i}^F\}_{i=1}^{n_R^F}$ and $\{Y_{R,i}^G\}_{i=1}^{n_R^G}$  was used for the estimation of the ROC curve (note that the samples of covariates $\{X_{R,i}^F\}_{i=1}^{n_R^F}$ and $\{X_{R,i}^G\}_{i=1}^{n_R^G}$ are not needed to compute the ROC curve), and the other, $\{(X_{A,i}^F,Y_{A,i}^F)\}_{i=1}^{n_A^F}$ and $\{(X_{A,i}^G,Y_{A,i}^G)\}_{i=1}^{n_{A^G}}$, for the estimation of the AROC curve, with $n^F=n_R^F+n_A^F$ and $n^G=n_R^G+n_A^G$. 

In order to approximate their distributions we are going to use a bootstrap algorithm. 
%The first challenge that we will face is that
However, the usual bootstrap methods are not directly applicable because replicating the null hypothesis in a study with ROC curves is not a straightforward matter. 
This is why the bootstrap algorithm that we will be using is based on the general bootstrap algorithm proposed in \cite{Martinez-Camblor2012}, aimed for problems with a complex data structure. 
First, we consider the expression
\begin{eqnarray}
T_{\psi} &=&  \psi \left( \left(\widehat{ROC}(p) - ROC(p) \right) -\left( \widehat{AROC}(p) - AROC(p) \right)  \right),\notag 
\end{eqnarray}
where $ROC$ and $AROC$ represent the theoretical ROC and AROC curves, respectively.  
Note that, under the null hypothesis, $ROC(p)=AROC(p)$ for $p\in (0,1)$, and therefore $S_{\psi} = T_{\psi}$ (the same applies for $\psi_{L_1}$, $\psi_{L_2}$, $\psi_{KS}$ or any other distance considered). The idea behind this methodology is to use $T_{\psi}$ instead of $S_{\psi}$ to compute the bootstrap statistic in the algorithm (note that, whereas $T_{\psi}$ cannot be calculated in practice, it can be calculated in a bootstrap environment). This way, we do not need to assume any hypothesis when generating the bootstrap samples: the null hypothesis is being used when we exchange $S_{\psi}^*$ by $T_{\psi}^*$. This idea has been already used in the context of ROC curves in \cite{Martinez-Camblor2013a},   \cite{Fanjul-Hevia2019} or \cite{Fanjul-Hevia2020}.

In \cite{Martinez-Camblor2012} the general bootstrap algorithm was designed for the comparison of a certain parameter or function in different populations, and here the ROC and the AROC curves that we want to compare come from the same place.  
To avoid the dependency problems that this entails, we will randomly split the original samples conformed by $\{(X_i^F,Y_i^F)\}_{i=1}^{n^F}$ and $\{(X_i^G,Y_i^G)\}_{i=1}^{n^G}$ in two sets. One of those, $\{Y_{R,i}^F\}_{i=1}^{n_R^F}$ and $\{Y_{R,i}^G\}_{i=1}^{n_R^G}$,  will be used for the estimation of the pooled ROC curve (note that the covariates are not needed to compute the ROC curve), and the other, $\{(X_{A,i}^F,Y_{A,i}^F)\}_{i=1}^{n_A^F}$ and $\{(X_{A,i}^G,Y_{A,i}^G)\}_{i=1}^{n_A^G}$, for the estimation of the AROC curve, with $n^F=n_R^F+n_A^F$ and $n^G=n_R^G+n_A^G$. 

This splitting of the sample does not have to be necessarily even. Given the greater complexity of the AROC curve it seems reasonable to employ more data for its estimation. Different partitions of the data will be further explored in the simulation study in Section \ref{sec:3}.

The proposed bootstrap algorithm to approximate the distribution of \eqref{eq:Sphi} goes as follows:
\begin{enumerate}
\item From the original sample, compute the statistic value $s_{\psi}$, using $\{Y_{R,i}^F\}_{i=1}^{n_R^F}$ and $\{Y_{R,i}^G\}_{i=1}^{n_R^G}$ for the estimation of the ROC curve and {\small{$\{(X_{A,i}^F,Y_{A,i}^F)\}_{i=1}^{n_A^F}$}} and {\small{$\{(X_{A,i}^G,Y_{A,i}^G)\}_{i=1}^{n_A^G}$}} for the estimation of the AROC curve as in \eqref{eq:AROCestim}

\item Generate $B$ random samples (with $B$ large) for the two sets of data. For $b\in\{1,\cdots,B\}$:
\begin{itemize}
\item[(i)] For $D\in \{F,G \}$, let $\{Y_{R,i}^{D,b*}\}_{i=1}^{n_R^D}$ be an i.i.d.~sample from the empirical distribution function obtained from the first set of data.
\item[(ii)] For $D\in \{F,G \}$, let $\{\varepsilon_{A,i}^{D,b*}\}_{i=1}^{n_A^D}$ be an i.i.d.~sample from the empirical distribution function of the residuals computed using the second set of data. Build the bootstrap sample $\{(X_{A,i}^D,Y_{A,i}^{D,b*})\}_{i=1}^{n_A^D}$,  where $Y_{A,i}^{D,b*} = \hat{\mu}^D(X_{A,i}^D)+\hat{\sigma}^D(X_{A,i}^D)\varepsilon_{A,i}^{D,b*}$.
\end{itemize}
\item  For $b\in\{1,\cdots,B\}$, obtain $\widehat{ROC}^{b*}(p)$ for $p\in (0,1)$ from $\{Y_{R,i}^{D,b*}\}_{i=1}^{n_R^D}$ (with $D\in \{F,G\}$) and $\widehat{AROC}^{b*}(p)$ for $p\in (0,1)$ from $\{(X_{A,i}^D,Y_{A,i}^{D,b*})\}_{i=1}^{n_A^D}$ (with $D\in \{F,G\}$).

\item  Using $T_\psi$ instead of $S_\psi$, compute the statistic bootstrap values $t_{\Psi}^{b,*}$, replacing $\widehat{ROC}$ by $\widehat{ROC}^{b*}$ , $ROC$ by $\widehat{ROC}$, $\widehat{AROC}$ by $\widehat{AROC}^{b*}$ , and $AROC$ by $\widehat{AROC}$ for $b\in \{1, \dots, B\}$.

\item Use, as a p-value approximation, $ p-value = \frac{1}{B} \sum_{b=1}^B I(s_\psi \leq t_{\psi}^{b,*})$.
\end{enumerate}

%=================================================================
\section{Simulations}
\label{sec:3}
%=================================================================

\begin{figure}[!tb]
\begin{center}
\includegraphics[scale=0.83]{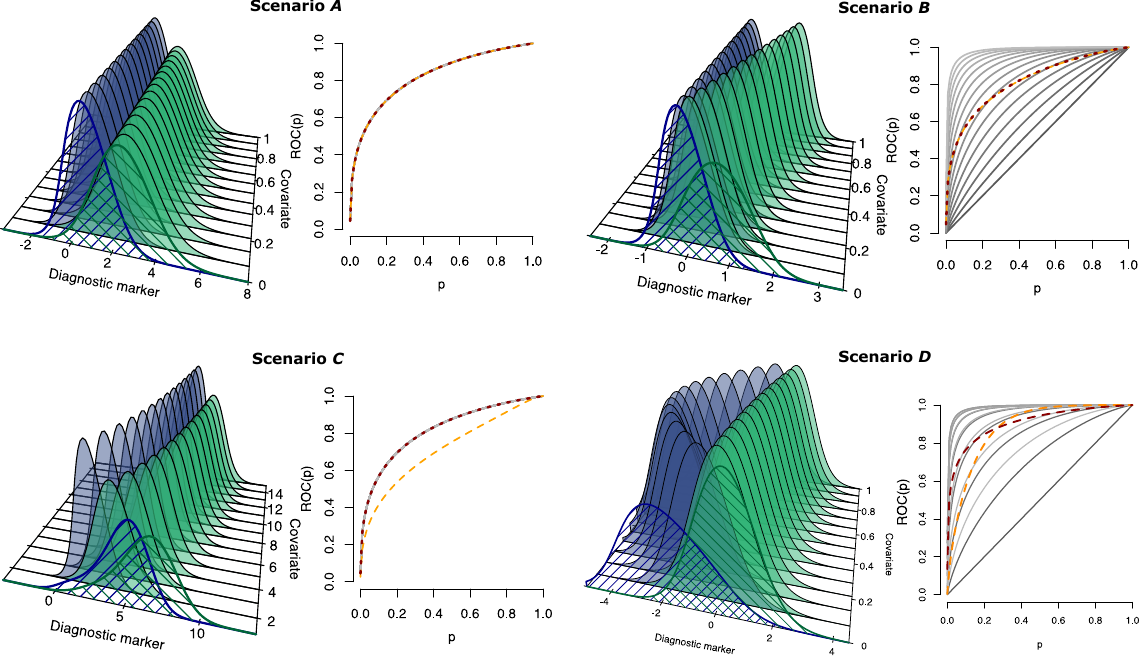}
\end{center}
\caption{Four scenarios with different relationships between the pooled, conditional and covariate-adjusted ROC curves. For each one of them, the conditional densities of the diagnostic variables (blue for the healthy population and green for the diseased population) are depicted for several fixed values of the covariate. The striped densities in the front row represent the densities for the marginal diagnostic markers. The corresponding ROC  and AROC curves are drawn for each case in a discontinuous orange and brown line, respectively. The conditional ROC curves are represented for every value of the covariate whose conditional densities are also represented.}
\label{fig:ScenariosandDenx}
\end{figure}

In this section we carry out a finite sample study to analyse the performance of this new test in terms of level approximation and power. We consider four different scenarios: $A$, $B$, $C$ and $D$,  represented in Figure~\ref{fig:ScenariosandDenx}. 

Each one of these scenarios is affected differently by a covariate, producing diverse relationships between the pooled (in orange), conditional (in gray) and AROC curves (in brown). Along with those curves, in  Figure~\ref{fig:ScenariosandDenx} we have drawn the densities of the diagnostic variables for the diseased (in green) and the healthy (in blue) populations. Apart from representing the densities of the pooled diagnostic samples (the ones with the striped areas) we selected several values of the covariate and draw their corresponding conditional densities (those same values are the same selected for the representation of the conditional ROC curves).

It is precisely the representation of those conditional densities what indicates when the covariate has an effect on the diagnostic variables. Scenario $A$ is the only one in which the conditional densities remain the same regardless the value of the covariate (and coincide with the marginal densities as well). In this case, it is obvious that the pooled, conditional and covariate-adjusted ROC curves have to be equal.

In scenario $B$, the situation changes: although the distribution of the diagnostic variable in the healthy population remains unchanged by the covariate, the same does not apply to the diseased population. Thus, for lower values of the covariate, the conditional densities overlap with each other almost completely, whereas this overlapping is reduced when the covariate increases, separating the conditional densities. This translates into conditional ROC curves that are very close to the diagonal for the lower values of the covariate and very close to the point of maximum specificity and sensitivity for the highest values of the covariate. The pooled ROC and the AROC curves coincide in this situation. 

Scenario $C$ shows a more curious situation: we have diagnostic variables that are affected by the covariate, but this effect is such that the discriminatory capability remains constant throughout all the values of the covariate. This means that the conditional ROC curves are equal, and that they match the AROC curve. In fact, if we calculate (following the models on Table~\ref{table:4scenarios} on page~\pageref{table:4scenarios}) the expression of this conditional ROC curve, for a certain $x\in R_X$, we obtain $ROC^x(p) = 1- \Phi\left(\frac{10}{13} \left(\Phi^{-1}(1-p) -\frac{3}{2}\right) \right)$ for $p \in(0,1)$, which is independent of the value of $x$. However, the effect of the covariate is noted when representing the pooled ROC curve, as it is attenuated with respect to the other two curves. In a practical situation this means that if we disregard the effect of this particular covariate, the performance of the diagnostic method would be compromised.

The last scenario, $D$, shows a situation in which the three curves are different. This time, both the lower and the higher values of the covariates produce conditional ROC curves close to the diagonal, whereas the medium values procure a wider separation of the corresponding conditional densities.

In these examples (and particularly in scenarios $B$ and $D$) we can  observe the interpretation of the AROC curve as a vertical average of the conditional ROC curves at each $p$. Note, however, that this average has to take into consideration the distribution of the covariate, which is not reflected in any way in Figure~\ref{fig:ScenariosandDenx} (the covariate values chosen to condition the densities and the ROC curves where selected uniformly on the support of the covariate).

It may not have a lot of sense to test the null hypothesis expressed in \eqref{eq:TestH02} on scenarios like $B$ or $D$, since we have already established  that their conditional ROC curve changes with the value of the covariate (and thus, that should be the curve considered for further studies). However, we have kept them in our simulation study  to show that this test does not need any assumption regarding the behaviour of the conditional ROC curve.

Note that there are some sufficient conditions that ensure the equality of certain curves:  when the diagnostic variable $Y^G$ is independent of the covariate (like in Scenarios A and B), the pooled ROC curve always coincides with the AROC curve. However, that is not a necessary condition. For example, a diagnostic variable with the same distribution for the diseased and the healthy population (even if such distribution depends on the value of the covariates) will result in the same pooled ROC, conditional and AROC curves: the diagonal of the one unit square.

The location-scale regression models  assumed for the construction of these four scenarios, similar to the one presented in \eqref{eq:locscale}, are specified in Table~\ref{table:4scenarios}. In that table we also indicate the relationships among the conditional, the AROC and the pooled ROC curves.

\begin{table}[!tb]
  \centering
%  \begin{spacing}{1.1}
\caption{Conditional mean and conditional standard deviation functions considered for the construction of Scenarios $A,B,C$ and $D$.}
\label{table:4scenarios}
\begin{tabular}{ m{1.48cm} m{4.4cm} m{3.8cm} m{4.65cm} }
  \hline  
 \thead{\textbf{Scenario}} & \thead{\textbf{Regression functions}}& \thead{\textbf{Conditional standard }\\\textbf{deviation functions}} &\thead{\textbf{Relationship}\\\textbf{among the curves}}  \\
  \hline
  \vspace*{3pt}
  \centering $ A$
  	&
  	\vspace*{3pt}
		$\mu_A^F ({x})= 2.5 $ 
		
		$\mu_A^G ({x})=  1 $
	&
\vspace*{3pt}
		$\sigma_A^F (x)=  1.3$ 
		
		$\sigma_A^G (x)= 1$
&
\vspace*{3pt}
 $ROC^x=AROC \; \forall x\in R_X$
 
  $ROC=AROC$
 	\\
\vspace*{5pt}
\centering  $ B$
  	&
  	\vspace*{5pt}
		$\mu_B^F ({x})=  1.5 x$
		
		$\mu_B^G ({x})=  0$
	&
\vspace*{5pt}
		$\sigma_B^F ({x})=  0.5$
		
		$\sigma_B^G ({x})=  0.5$
&
\vspace*{5pt}
$ROC^x \neq AROC \;   x\in R_X$
 
  $ROC=AROC$
	\\
	\vspace*{5pt}
 \centering  $C $
  	&
  	\vspace*{5pt}
		$\mu_C^F ({x})= 2.5+2\log(x)  $
		
		$\mu_C^G ({x})= 1+2\log(x) $
	&
\vspace*{5pt}
		$\sigma_C^F ({x})=  1.3$
		
		$\sigma_C^G ({x})=  1$
&
\vspace*{5pt}
$ROC^x=AROC \; \forall x\in R_X$
 
  $ROC \neq AROC$
 	\\  
 	\vspace*{5pt}
 \centering  $D $
  	&
  	\vspace*{5pt}
		$\mu_D^F (x)= x^2 $
		
		$\mu_D^G (x)=  3\sin(\pi (x+1))$
	&
\vspace*{5pt}
		$\sigma_D^F (x)=  1$
		
		$\sigma_D^G (x)=  1$
&

\vspace*{5pt}
$ROC^x\neq AROC \;  x\in R_X$
 
  $ROC\neq AROC$
\\
  \hline
\end{tabular}
%\end{spacing}
\end{table}

 The regression errors $\varepsilon^F$ and $\varepsilon^G$ were considered to follow  normal standard distributions. The covariate followed a uniform distribution on the unit interval for both the diseased and the healthy population in Scenarios $A,B$ and $D$. For Scenario $C$, the covariate follows a uniform distribution on the interval $[1,15]$.
 Three different sample sizes were considered for the study, with $(n^F,n^G) = (100,100), (250, 350), (500,500)$. Note that the second pair of sample sizes is unbalanced. 1000 datasets were simulated to compute the proportion of rejection for each case. The number of bootstrap iterations  considered was $B=200$.
 
 We used the three different distance functions previously mentioned  ($\psi_{L_1}$, $\psi_{L_2}$  and  $\psi_{KS}$) for the construction of the test statistic, so we would discuss the results for the three of them. We will be denoting them as the $L_1$ (the one based on the $L_1$ measure), the $L_2$ (the one based on the $L_2$ measure) and the $KS$ (the one based on Kolmogorov-Smirnov criterion) statistics.

 Three different partitions of the sample for the estimation of the ROC and the AROC curves (needed to avoid dependency issues) were considered: in the first one, the splitting of the sample was even, meaning $n_R^F = n_A^F = 1/2 n^F$ and $n_R^G = n_A^G = 1/2 n^G$; in the second case, one third of the sample was used for the estimation of the pooled ROC curve, whereas the remaining 2/3 was used for the AROC curve (i.e., $n_R^F = 1/3n^F$,  $n_A^F = 2/3 n^F$, $n_R^G =1/3 n^G$ and $n_A^G = 2/3 n^G$); in the last case, the proportion was 1/4 for the ROC curve and 3/4 for the AROC curve (i.e., $n_R^F = 1/4n^F$,  $n_A^F = 3/4 n^F$, $n_R^G =1/4 n^G$ and $n_A^G = 3/4 n^G$). 

Scenarios $A$ and $B$ were the ones selected to calibrate the level of the test (as they have equal ROC and AROC curves, they meet the null hypothesis). We show the results for three different nominal levels: $\alpha\in \{0.025,0.05, 0.1\}$. Scenarios $C$ and $D$ were used to analyse  the power. Note that the separation between the two curves is wider in scenario $C$, so we expect to obtain a higher power there with respect to  scenario $D$. For those last scenarios we only show the results for $\alpha=0.05$. 

\begin{figure}[!tb]
\begin{center}
\includegraphics[scale=0.62]{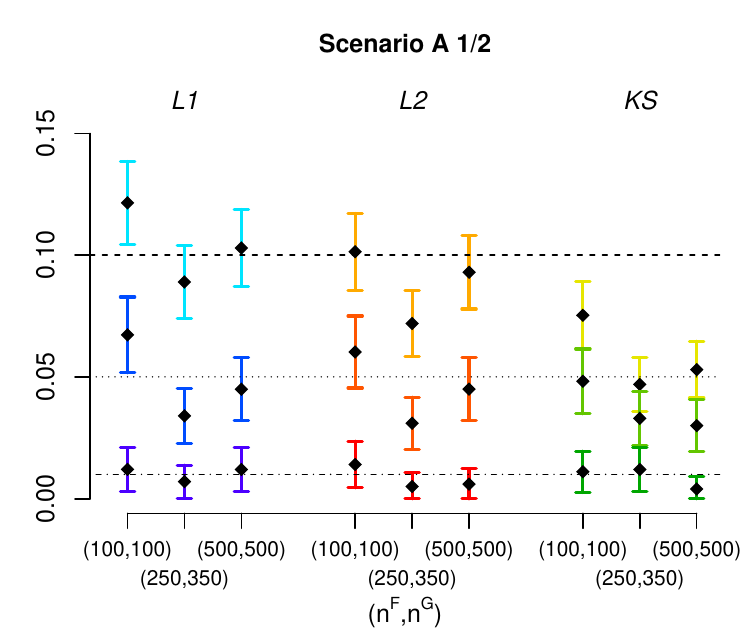}
\includegraphics[scale=0.62]
{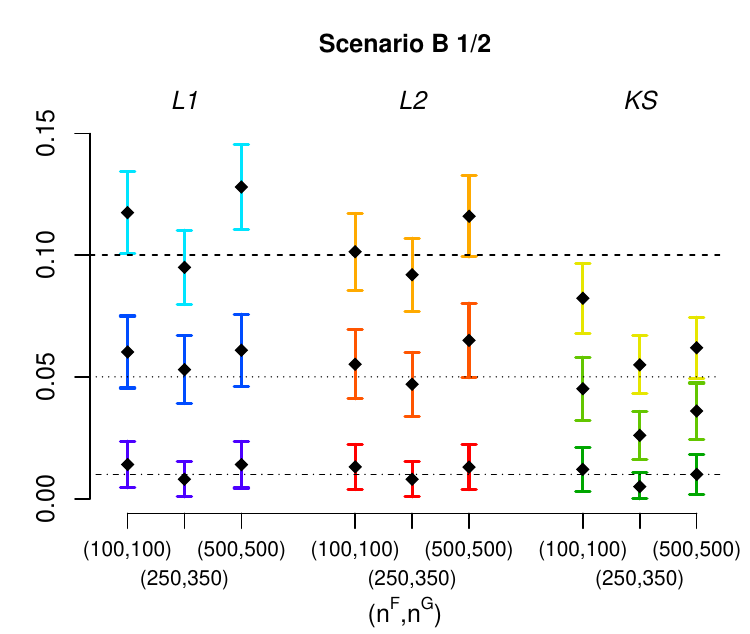}

\includegraphics[scale=0.62]{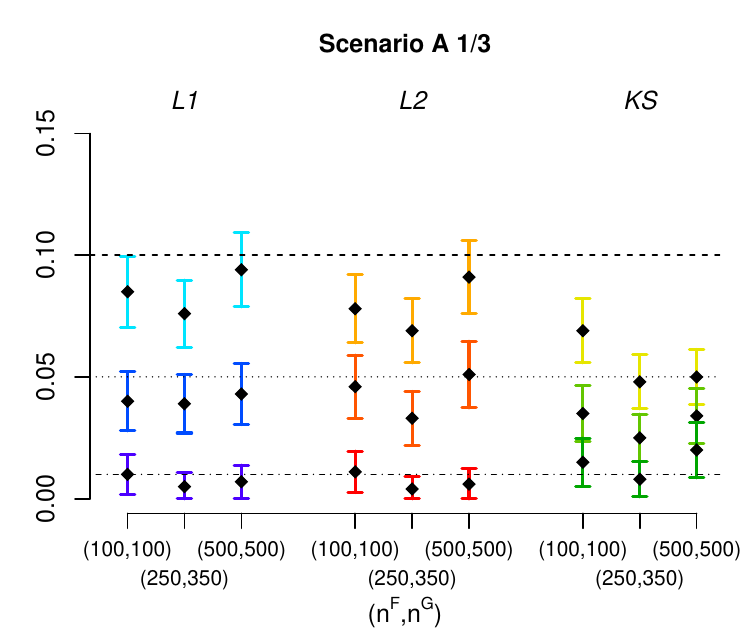}
\includegraphics[scale=0.62]
{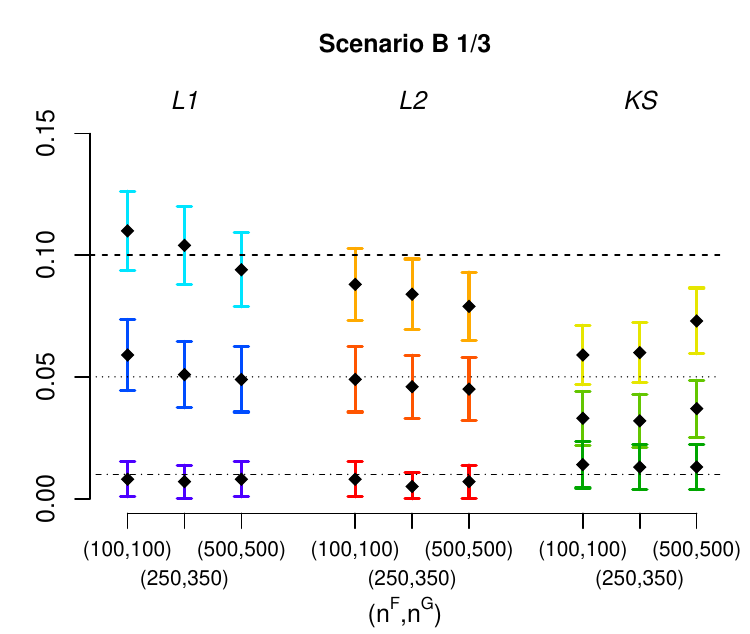}

\includegraphics[scale=0.62]{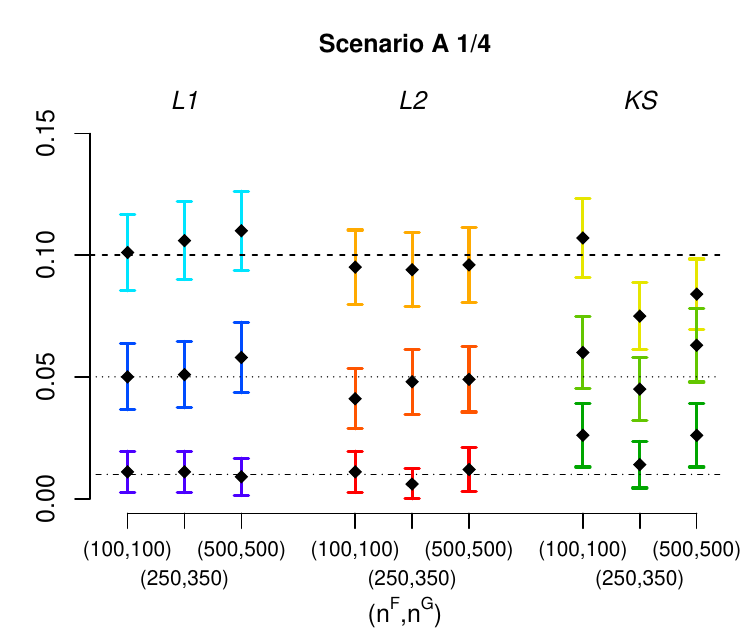}
\includegraphics[scale=0.62]
{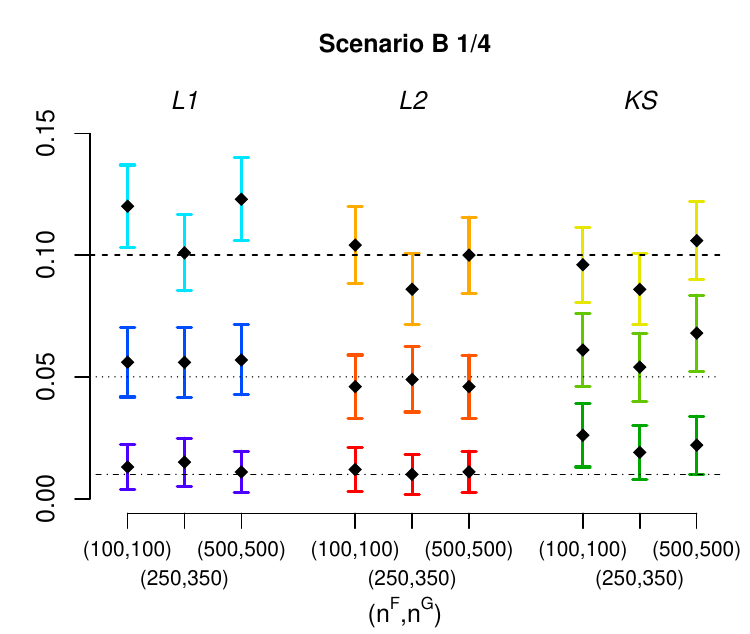}
\vskip-10pt
\includegraphics[scale=0.55]{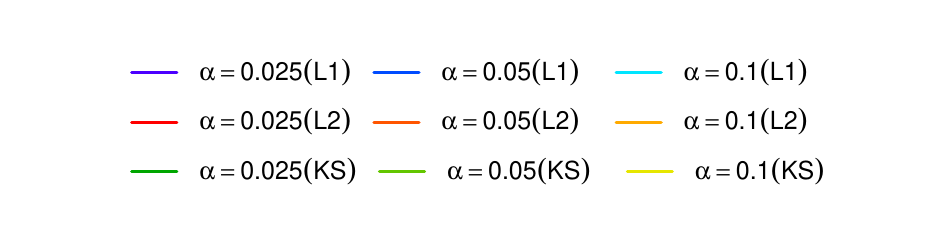}
%\vskip-10pt
\end{center}
\vskip-10pt
\caption{Results of the simulations study for under the null hypothesis. The proportion of the sample that was used for the estimation of the ROC curve is indicated in the name of each graph to differentiate the partitions considered. 
These graphs (one columns for each scenario under the null hypothesis) show the estimated proportions of rejection and their corresponding confidence intervals for all the sample sizes and the three test statistics considered.}
\label{fig:ResSimH0}
\end{figure}

In Figure~\ref{fig:ResSimH0} we have six graphs containing the results of the simulation study for scenarios A and B, each row  representing a different partition of the sample. The proportion of rejections are displayed there for each sample size and each test statistic.
 In order to obtain more detailed information about the practical performance of the tests under the null hypothesis we have included intervals constructed around the estimated proportion of rejection to verify whether the level is correctly approximated. More specifically, for a given estimated proportion of rejections, $\hat{p}$, the shown interval is $\left[\hat{p} \pm 1.96 \sqrt{\frac{\alpha(1-\alpha)}{n_s}}\right]$, where $n_s$ is the number of simulated samples used to obtain the estimated proportion. 
 As long as those intervals contain the nominal level we can say that the test is well calibrated, as this is equivalent to perform a test to check if the actual level of the test equals the nominal level $\alpha$. Note that the sample $n_s$ is, in this case,  1000 for all intervals considered, and thus their length is not influenced by the sample sizes of the ROC curves of the study. %(strictly speaking they are not confidence intervals, but we will use this notation for the sake of simplicity).

In the case of the $L_1$ and $L_2$ statistics,  the nominal levels are in general well approximated by the estimated proportions, although in Scenario $B$ the result for the unbalanced sample size is a bit overestimated. The $KS$ statistic, however, seems to be more conservative (which is in line with the conservativeness of the Kolmogorov-Smirnov test).
 The calibration of the test improves when the partition of the sample benefits the estimation of the AROC curve (specially for the $KS$ statistic), although this improvement is less noticeable for the higher sample sizes considered (which, on the other hand, is to be expected).

\begin{table}[!t]
\centering
\caption{Results of the simulations study under the alternative hypothesis. The estimated proportions of rejection for scenarios $C$ and $D$ are given for the three pairs of sample sizes $(n^F,n^G)$ considered, for the three statistics that use different distance functions and for the three ways of splitting the sample to avoid dependency problems ($1/2$, $1/3$, $1/4$).}
\label{tab:ResSimH1}
\begin{tabular}{lrrrrlrrrlrrr}
\hline
    & \multicolumn{1}{c}{} & \multicolumn{3}{c}{\thead{Statistic $L_1$} }&  & \multicolumn{3}{c}{\thead{Statistic $L_2$} }&  & \multicolumn{3}{c}{\thead{Statistic $KS$}} \\
    \cline{3-5} \cline{7-9} \cline{11-13} 
    & \thead{$(n^F,n^G)$}           & 1/2                & 1/3                & 1/4               &  & 1/2                & 1/3                & 1/4               &  & 1/2                & 1/3                & 1/4               \\
\hline
\thead{Scenario $C$} & $(100,100)$          & 0.272              & 0.228              & 0.163             &  & 0.283              & 0.220              & 0.159             &  & 0.211              & 0.130              & 0.099             \\
                                      & $(250,350)$          & 0.629              & 0.581              & 0.464             &  & 0.631              & 0.575              & 0.453             &  & 0.290              & 0.260              & 0.229             \\
                                      & $(500,500)$          & 0.852              & 0.808              & 0.712             &  & 0.844              & 0.802              & 0.705             &  & 0.433              & 0.358              & 0.314             \\ 
\hline
\thead{Scenario $D$} & $(100,100)$          & 0.142              & 0.086              & 0.081             &  & 0.223              & 0.101              & 0.074             &  & 0.369              & 0.166              & 0.118             \\
                                      & $(250,350)$          & 0.486              & 0.361              & 0.323             &  & 0.630              & 0.420              & 0.317             &  & 0.685              & 0.506              & 0.388             \\
                                      & $(500,500)$          & 0.832              & 0.715              & 0.580             &  & 0.884              & 0.725              & 0.548             &  & 0.839              & 0.697              & 0.580             \\ 
\hline
\end{tabular}
\end{table}

\begin{figure}[!tb]
\begin{center}
\includegraphics[scale=0.6]{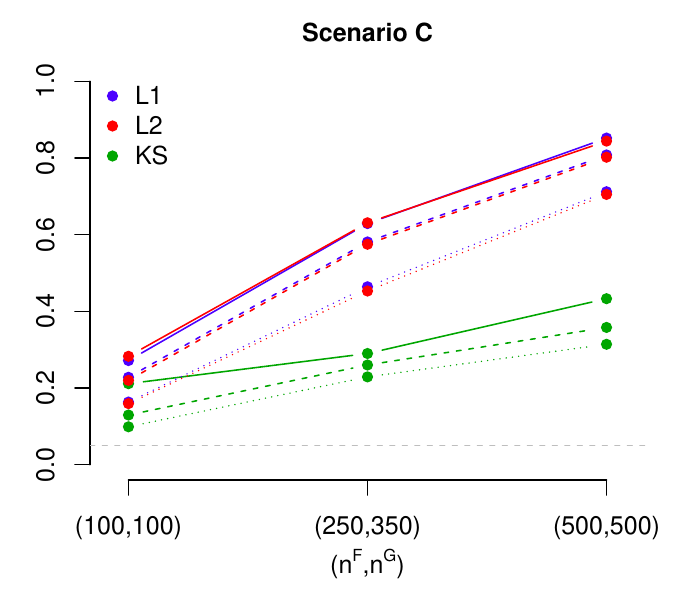}\qquad
\includegraphics[scale=0.6]{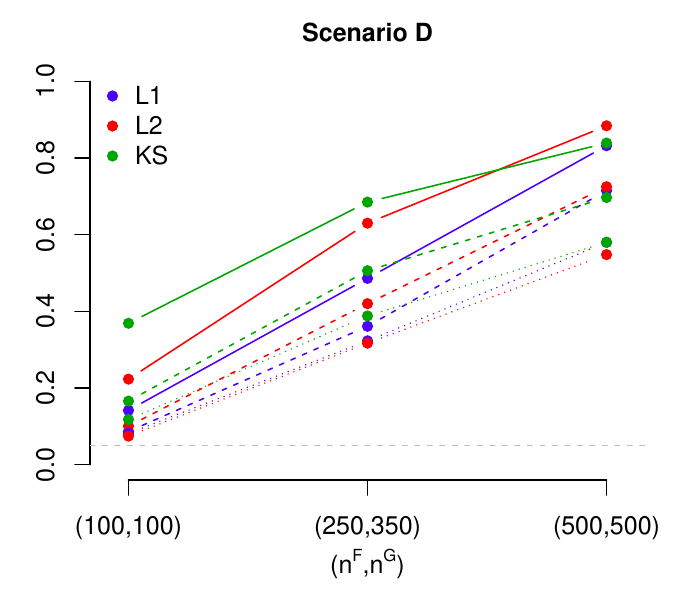}
\end{center}
%\vskip-10pt
\caption{Results of the simulations study under the alternative hypothesis. Each dot represents the  estimated proportions of rejection for each scenario, each partition of the sample (the solid line represent the 1/2 partition, the dashed lines represent the 1/3 partition and the doted lines the 1/4 partition), each sample sizes and each test statistics considered.}
\label{fig:ResSimH1}
\end{figure}

In Table~\ref{tab:ResSimH1} and Figure~\ref{fig:ResSimH1} the results of the simulation study regarding the power of the test are depicted, reflecting the consistency of the test: the power grows when we increase the sample size for all statistics. It is higher on Scenario $C$, as the difference between the ROC and the AROC curves is greater. 
In the situations where the partition of the sample for the estimation of the ROC and the AROC curves is more uneven, the power is also lower.

It is also worth noticing that, given that the ROC and the AROC curves of scenario D cross each other, a test statistic based on comparing the corresponding areas under the curve \citep[as it is commonly used in the literature when comparing ROC curves, like in][]{DeLong1988} could have no power to discriminate these curves.

%=================================================================
\section{Application}
\label{sec:4}
%=================================================================

In order to illustrate the discussion and the test developed in this paper we analyse a dataset concerning patients suspected of prediabetes provided by Dr. F. Gude Sampedro (Unidade de Epidemioloxía Clínica, Hospital Clínico Universitario de Santiago). 
Here, a patient is considered as prediabetic when it presents a diagnostic of diabetes mellitus or blood glucose levels above 100 mg/dl. Of the 1496 patients contained in this data set, 405 were considered as diseased (prediabetic) and 1091 as healthy. Note that this means that the sample sizes are unbalanced. 

Apart from the binary output that indicates if a patient has prediabetes or not, there are other variables in the dataset. On the one hand, we have three variables that we are going to consider as three different diagnostic markers: \textit{GA} (which represents the glycated albumin), \textit{a1c1} (haemoglobin) and  \textit{$-$GP22} (glycan peaks).  On the other hand, we are also going to take into account one covariate, the \textit{age}. In Table~\ref{table:Diabetes2} a summary of the continuous variables that are being used is shown. Our objective is to assess the capability of those three diagnostic variables to correctly diagnose  prediabetes while taking into account the covariate \textit{age}. In order to do that we have to determine if we should use the pooled ROC, the AROC or the conditional ROC curve for each diagnostic marker.

\begin{table}[!tb]
\centering
\caption{Summary of the variables contained in the \textit{Diabetes} dataset for the prediabetic (D) and the non-prediabetic (H) subjects.}
\label{table:Diabetes2}
\begin{tabular}{cccccccccccc}
    \hline
 &  \multicolumn{2}{c}{\thead{\textit{GA}}} &  & \multicolumn{2}{c}{\thead{\textit{a1c1}}}  & &\multicolumn{2}{c}{\thead{\textit{$-$GP22}}} & & \multicolumn{2}{c}{\thead{\textit{age}}} \\ 
 \cline{2-3}
  \cline{5-6}
   \cline{8-9}
    \cline{11-12}
	   & D    & H     & & D    & H    & & D    & H      & &  D    & H   \\
  \hline
Minimum    &  8.50&  7.88 & & 4.80 & 3.10 & & $-9.25$ & $-11.29$ & & 24.0 &18.00\\
1st quartile  & 13.46& 12.46 & & 5.60 & 5.20 & & $-6.11$ &  $-7.24$ & & 55.0 &35.00\\
Median & 15.11& 13.55 & & 5.90 & 5.30 & & $-5.43$ &  $-6.27$ & & 65.0 &47.00\\
Mean   & 15.96& 13.54 & & 6.31 & 5.35 & & $-5.53$ &  $-6.39$ & & 63.6 &48.56\\
3rd quartile  & 17.63& 14.58 & & 6.70 & 5.50 & & $-4.81$ &  $-5.46$ & & 73.0 &62.00\\
Maximum    & 33.96& 20.29 & & 12.80& 6.90  && $-3.23$ &  $-1.03$ & & 90.0 &91.00\\
   \hline
\end{tabular}
\end{table}

\begin{figure}[!tb]
\begin{center}
\includegraphics[scale=0.48]{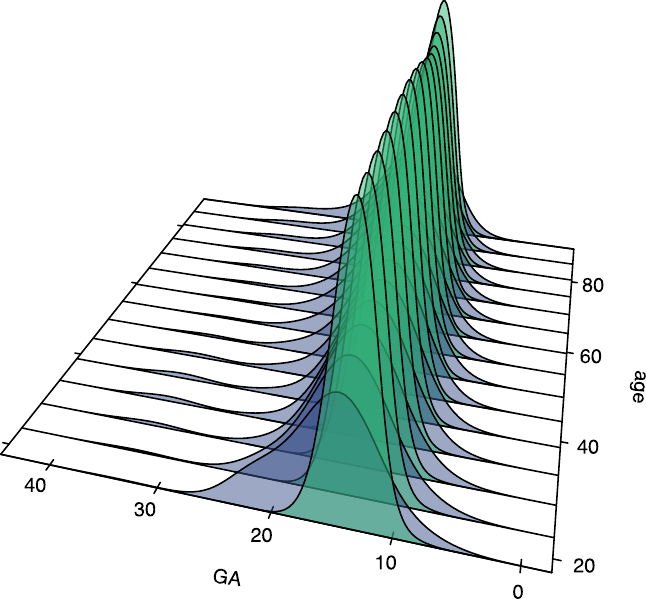}
\includegraphics[scale=0.48]{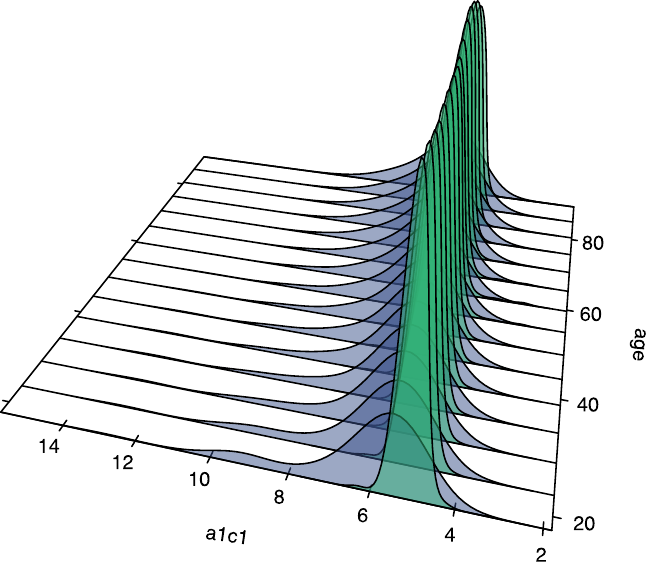}
\includegraphics[scale=0.48]{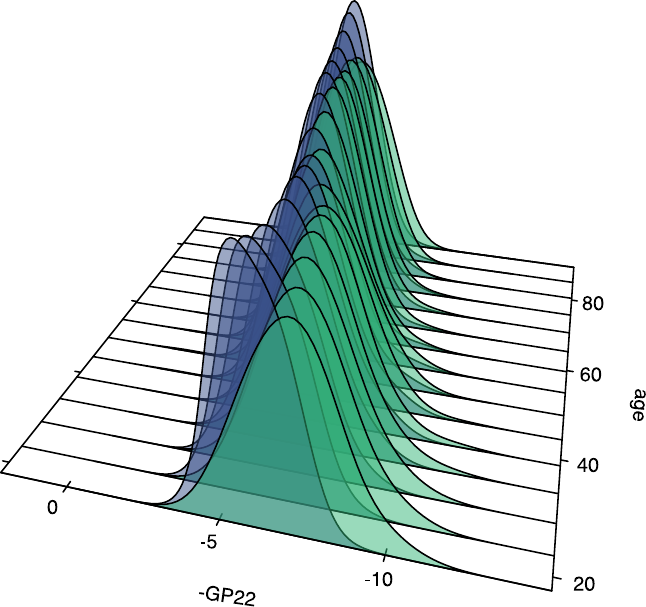}

\vskip6pt
\includegraphics[scale=0.45]{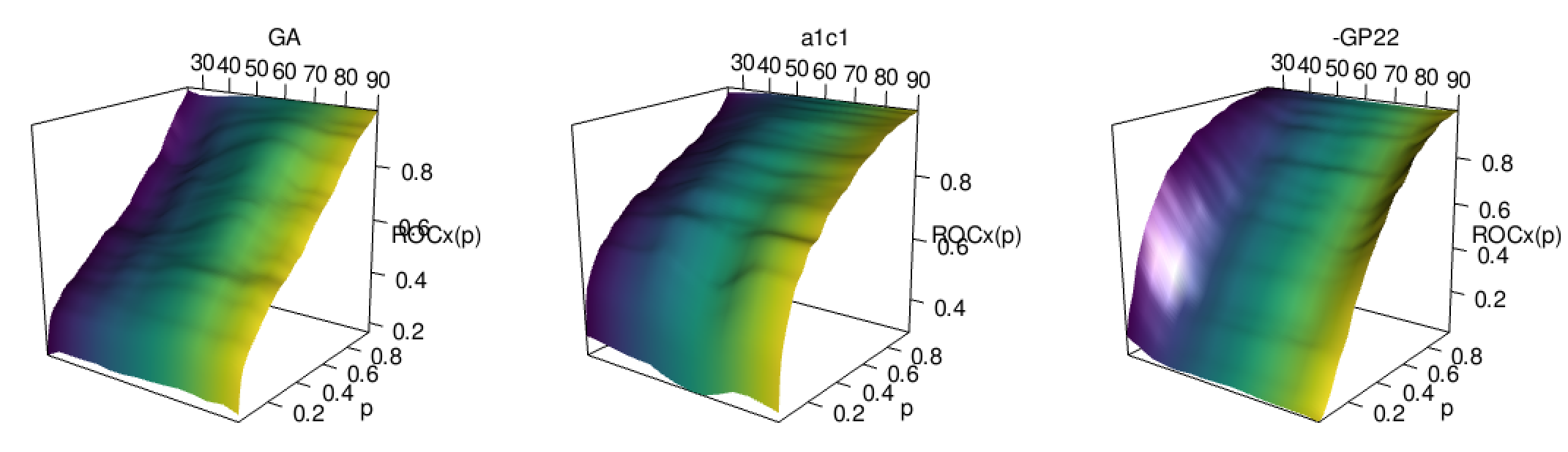}
\end{center}
\caption{Estimated conditional densities of the three diagnostic variables, taking \textit{age} as the continuous covariate and their corresponding estimated conditional ROC curves.}
\label{fig:AplDensitiesROCx}
\end{figure}

We begin our analysis by representing the conditional densities of the three diagnostic markers at certain ages, along with their corresponding conditional ROC curves. The resulting graphics are collected in Figure~\ref{fig:AplDensitiesROCx}. Note that the third diagnostic variable appears now under the tab \textit{$-$GP22}. This is because, in this particular case, higher values of the diagnostic variables are more common in the healthy population, whereas the diseased subjects tend to have lower values, which goes against the assumptions made for the construction of a ROC curve. By taking the opposite values of this variable we ensure that the roles are exchanged. 

 At first sight it could appear that the conditional ROC curves remain constant through all those values, although we can appreciate a sort of hill for the medium age in the \textit{GA} marker, and the \textit{$-$GP22} seems to have better discriminatory power for the youngest patients, as the conditional ROC curves at those lower ages are closer to the point of maximum sensitivity and specificity. 

However, there are two different issues that must be taken into consideration. First,  the conditional ROC curve is estimated locally, which means that the estimations computed on the extreme values of the covariate are not as reliable, because they have fewer data around (and this condition exaggerates when the covariate is not uniformly distributed). Secondly, on those representations there is no insight on how the covariate is distributed in the healthy and in the diseased populations.

\begin{figure}[!tb]
\begin{center}
\includegraphics[scale=0.95]{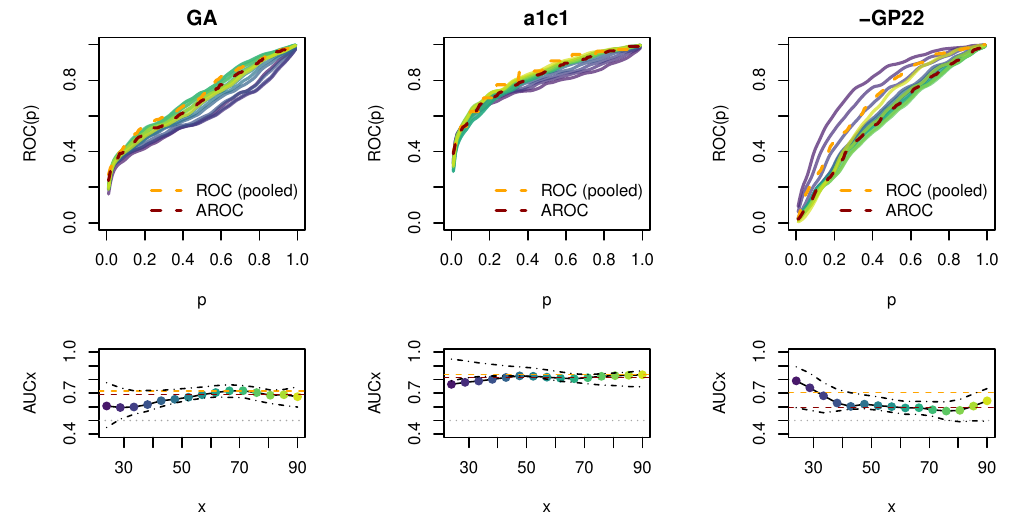}
\end{center}
\caption{Estimated pooled (in discontinuous orange lines), covariate-adjusted (in  discontinuous brown lines) and conditional (in continuous lines, one colour for each conditioned value) ROC curves of the three diagnostic, along with their corresponding summary measures, AUC (in discontinuous orange lines), AAUC (in discontinuous brown lines) and $AUC^x$, with its pointwise confidence interval. The gray horizontal line represents an AUC of 0.5, the hazard.}
\label{fig:ApROCx2D}
\end{figure}

Next, we estimated the pooled and the covariate-adjusted ROC curves for each one of the diagnostic variables. We represented them in Figure~\ref{fig:ApROCx2D}. The conditional ROC curve was also estimated for certain values of the covariate, as well as their respective conditional area under the curve (AUC) with a pointwise 0.95 confidence interval \citep[for more details of how to compute such confidence interval, check][]{Gonzalez-Manteiga2011a}. The summary measures AUC and area under the AROC curve (AAUC) were also estimated (they are represented as horizontal lines, as they do not depend on fixed values of the covariate). 

Setting our attention on those summary measures and the pointwise confidence interval we can have a first insight of the relationship between the curves. For the considered confidence level, the AAUC falls inside the confidence interval for all the values of the covariate, for all the diagnostic markers. Of course, we have to take into account that it is not a confidence band, so the level should be adjusted, but in any case it seems that there may not be differences between those indices. The AUC and the AAUC, despite being presented without confidence intervals, seem to be very similar in the first two variables. The ROC and the AROC curves of \textit{$-$GP22}, however, are more separated (as their corresponding summary measures are).

%\cite{ManRodriguez-Alvarez2017}
Then, we perform a two-step study for 
 each one of those diagnostic markers: first we test if the conditional ROC curve is constant for all the values of the covariate by comparing it with the AROC curve \citep[using the test proposed in ][]{Rodriguez-Alvarez2011} and then we test whether the AROC and the ROC curves are equal \eqref{eq:TestH02} (using the test statistic $L_2$ described in \ref{sec:2}, doing an even splitting of the sample for the dependency issues and considering 500 bootstrap iterations).  
  The software \textit{R} \citep{ManR} was used to run the analysis: the package \textit{npROCRegression} \citep{ManRodriguez-Alvarez2017} was used for the first test, and the code implemented for the second test is provided the Appendix~\ref{sec:Ap}.
 %The software \textit{R} \citep{ManR} was used to run the analysis, and in particular de package \textit{npROCRegression} \citep{ManRodriguez-Alvarez2017} was used for the first test.
 The  obtained p-values, with their interpretation when we take a significance level of $\alpha = 0.05$, are summarized in Table~\ref{table:results1}.
  \begin{table}[!tb]
\centering
\caption{Summarized p-values for the two-step study for the three diagnostic markers, taking $\alpha = 0.05$.}
\label{table:results1}
\begin{tabular}{ccccccc}
\hline
 & & \thead{\textit{GA}} & & \thead{\textit{a1c1}} & &\thead{\textit{$-$GP22} } \\ 
\cline{3-3} 
\cline{5-5}
\cline{7-7}  
\textbf{Test 1} & & $p-value=0.073 $& &$p-value=0.853$ & &$p-value =  0.523$  \\ 
\textbf{Test 2} & &$p-value= 0.782$& &$p-value=0.910 $ & &$p-value < 0.001$  \\    
 &  & $\Downarrow$ & &$\Downarrow$ & &$\Downarrow$  \\ 
   
$\alpha = 0.05$ &  & Use {\color{orange} $\bm{ROC}$} & & Use {\color{orange} $\bm{ROC}$} &  & Use {\color{darkred} $\bm{AROC}$}  \\
  % & {\color{gray} $\bm{ROC^x}$}  & {\color{orange} $\bm{ROC}$} & {\color{darkred} $\bm{AROC}$}  \\
  \hline
\end{tabular}
\end{table}
The conclusions that are drawn from that study match our previous suspicions: the covariate \textit{age} does not seem to have a significant impact on the performance of each diagnostic marker. However, in the case of the \textit{$-$GP22} marker we find differences between the ROC and the AROC curve, and thus the latter should be employed for further analysis of this diagnostic variable.

For the sake of the illustration (although this is not something that should be done in a practical case)  we have reviewed the obtained results, this time for a significance level of $\alpha=0.1$. The results are summarized in Table~\ref{table:results2}.
\begin{table}[!tb]
\centering
 \caption{Summarized p-values for the two-step study for the three diagnostic markers, taking $\alpha = 0.1$.}
 \label{table:results2}
\begin{tabular}{ccccccc}
 \hline
 & & \thead{\textit{GA}} & & \thead{\textit{a1c1}} & &\thead{\textit{$-$GP22} } \\ 
\cline{3-3} 
\cline{5-5}
\cline{7-7}  
\textbf{Test 1} & &$p-value=0.073 $& &$p-value=0.853$ & &$p-value =  0.523$  \\ 
\textbf{Test 2} & &$p-value= 0.782$& &$p-value=0.910 $ & &$p-value < 0.001$  \\    
   & &$\Downarrow$ & &$\Downarrow$ & &$\Downarrow$  \\ 
   $\alpha=0.1$ & & Use {\color{gray} $\bm{ROC^x}$} & &Use {\color{orange} $\bm{ROC}$} & & Use {\color{darkred} $\bm{AROC}$}  \\
   \hline
   \end{tabular}
\end{table}
 The conclusions drawn this time are very similar for the diagnostic markers of \textit{a1c1} and \textit{$-$GP22}, but for the \textit{GA} variable the first test rejects the null hypothesis, indicating that its performance as a diagnostic method can change depending on the values of \textit{age}.

Another aspect that should be considered is that we are performing sequential comparisons without taking into account the problems that can arise from multitesting, but we do not elaborate further in this topic, as it is not the aim of this study. However, in a practical situation the level $\alpha$ should be controlled.

%=================================================================
\section{Discussion}
\label{sec:5}
%=================================================================

In this paper we have shown how a covariate can affect the performance of an ROC curve study in several ways. We have discussed the different curves (pooled ROC, AROC and conditional ROC curve) that can be used to incorporate the covariate effect in the analysis, and the scenarios that can arise depending on the existing relationships between the three curves. A new nonparametric test was proposed in that context for comparing the pooled ROC and the AROC curves for the case in which there is one continuous covariate. A bootstrap algorithm was proposed to approximate the statistic distribution and a simulation study was carried out reflecting the good behaviour of the proposed methodology.

Despite the fact that throughout this document we have been dealing with a unidimensional covariate, the discussion of how to asses the significance of its effect in an ROC curve study is still valid for a multidimensional covariate. The limitation of the new test proposed in Section~\ref{sec:2}  \citep[as well as the one proposed by][]{Rodriguez-Alvarez2011} comes mostly from the considered estimators of the conditional and covariate-adjusted ROC curves, which are valid only for unidimensional covariates. The proposed test statistic could be extended to a multidimensional covariate scenario as long as an adequate estimator for the AROC curve is provided and the resampling step in the bootstrap algorithm is adapted accordingly.  One alternative could be to use single-index models in order to reduce the dimension of the covariate to the unidimensional case (although this would entail introducing a semi-parametric component to the methodology).
%the bootstrap resampling step is adapted.

To avoid the dependency problems that would arise when computing the estimators of the ROC and the AROC curves from the same sample, the sample was split in two disjointed sets. This solution, although effective, may imply a loss of power. The search for an alternative procedure that overcomes this disadvantage is deferred for future studies. Still, to the best of our knowledge, this is the first methodology designed for this kind of test, with the upside of being a nonparametric approach. 

How to decide the proportion of the sample that is saved for the AROC curve estimation in such partition is another issue. The simulation study shows that, when the sample size is small, the test is better calibrated for an uneven partition that benefits the AROC curve estimation, but for higher sample sizes this is no necessarily true. Given that it also shows that the even partition is the one that yields higher power, our recommendation is to split the sample evenly as long as the sample size is big enough. The conclusions reached in the application with real data illustrated in Section~\ref{sec:4} do not change when considering a 1/2, a 1/3 or a 1/4 partition.

%The splitting of the sample that was ...to compute the estimators of the AROC and ROC curves implies a loss of power, as the sample size is divided by half. There is some room for improvement, and the search for an alternative procedure that overcomes this disadvantage is deferred for future studies. Still, to the best of our knowledge, this is the first methodology designed for this kind of test, with the upside of being a nonparametric approach. 

Most of the tests that exists in the literature for the comparison of ROC curves are based on the comparison of the areas under those curves. For our particular problem this would translate to base the test in the comparison of the AUC and the AAUC. The problem with this approach is that, even if we were able to prove that the summary measures were equal, this would not necessarily mean that their corresponding curves are equal (although the converse implication is true). The methodology considered here compares the whole curves involved and, thus, avoids this issue so we do not to worry if our curves cross each other.

It is also worth noticing that this methodology can be applied regardless on the effect that the covariate may have in the conditional ROC curve. Of course, when the shape of the conditional ROC curve changes with the value of the covariate, we would not be concerned with the pooled ROC curve (and by extension, with this test). However, in cases with scarce data where the estimator of the conditional ROC curve could not be reliable, this test could still be useful to assess the significance of the covariate.

Knowing when we need to use the conditional ROC curve, the AROC curve or if we can completely disregard the covariate information and use the pooled ROC curve is of the essence for the further analysis that we may be interested in. All the possible applications that the study of ROC curves may have (from searching for optimal cut-off points to comparing different diagnostic markers) are sensitive to the presence of covariates.

One of the motivations for correctly acknowledging the covariate effect in this context is the fact that the thresholds of the diagnostic variables can have different sensitivities and specificities for different covariate values. Thus, we could now look for optimal thresholds for the pooled, conditional or AROC curves. Moreover, if our objective were to compare diagnostic methods throughout the comparison of the corresponding ROC curves in the presence of covariate information, depending on how the considered covariate affects each marker, the type of ROC curve that should be used for the comparison of the markers can change.

%=================================================================
\appendix
\section{Appendix}
\label{sec:Ap}
%=================================================================

The code in R that allows to perform the test \eqref{eq:TestH02} using the methodology described in this document is available in the following GitHub repository: \newline \url{https://github.com/arisfanjul/CovEffectinROCcurves?tab=readme-ov-file#readme}

It includes a manual of the functions provided there as well an illustrative example. The number of bootstrap iterations and the way the sample is splitted can be adjusted. 
The output of the test gives the value of the test statistic defined in \eqref{eq:Sphi} for the three different distance functions here considered, and the corresponding p-values. The estimation of both the pooled ROC curve \eqref{eq:ROCestemp} and the AROC curve \eqref{eq:AROCestim} is also provided.

%%%%%%%%%%%%%%%%%%%%%%%%%%%%%%%%%%%%%%%%%%%%%%%%%%%%%%%%%%%
\section*{Acknowledgements}
%The authors would like to thank the Associate Editor and the anonymous reviewers for their constructive comments and suggestions on an earlier version of this manuscript. 

The research of A. Fanjul-Hevia,W. González-Manteiga   is supported by the Grant PID2020-116587GB-I00 from Spanish Ministerio de Ciencia e Innovación (MCIN/AEI/ 10.13039/501100011033). J.C. Pardo-Fernández acknowledges financial
support by the Grant PID2020-118101GB-I00 from Spanish Ministerio de Ciencia e Innovación (MCIN/AEI/10.13039/501100011033).
A. Fanjul-Hevia and J.C. Pardo-Fernández also acknowledge the support from Grant PID2023-148811NB-I00 from Spanish Ministerio de Ciencia, Innovación y Universidades (MICIU/AEI/10.13039/501100011033) and by European Union ERDF. 
Dr. F. Gude (Unidade de Epidemioloxía Clínica, Hospital Clínico Universitario de Santiago) is thanked for providing the data set analysed in this article.

% ORCID
%A. Fanjul-Hevia http://orcid.org/0000-0002-3254-8711
%J. C. Pardo-Fernández http://orcid.org/0000-0002-8260-374X
%W. González-Manteiga http://orcid.org/0000-0002-3555-4623

%%%%%%%%%%%%%%%%%%%%%%%%%%%%%%%%%%%%%%%%%%%%%%%%%%%%%%%%%%%
% Biblio
%%%%%%%%%%%%%%%%%%%%%%%%%%%%%%%%%%%%%%%%%%%%%%%%%%%%%%%%%%%

\bibliographystyle{apalike}

\bibliography{biblio}

\begin{thebibliography}{}

\bibitem[DeLong et~al., 1988]{DeLong1988}
DeLong, E.~R., DeLong, D.~M., and Clarke-Pearson, D.~L. (1988).
\newblock Comparing the areas under two or more correlated receiver operating
  characteristic curves: a nonparametric approach.
\newblock {\em Biometrics}, 44:837--845.

\bibitem[Fanjul-Hevia et~al., 2021]{Fanjul-Hevia2019}
Fanjul-Hevia, A., Gonz{\'a}lez-Manteiga, W., and Pardo-Fern{\'a}ndez, J.~C.
  (2021).
\newblock A non-parametric test for comparing conditional {ROC} curves.
\newblock {\em Computational Statistics \& Data Analysis}, 157:107146.

\bibitem[Fanjul-Hevia et~al., 2024]{Fanjul-Hevia2020}
Fanjul-Hevia, A., Pardo-Fern{\'a}ndez, J.~C., Van~Keilegom, I., and
  Gonz{\'a}lez-Manteiga, W. (2024).
\newblock A test for comparing conditional {ROC} curves with multidimensional
  covariates.
\newblock {\em Journal of Applied Statistics}, 51(1):87--113.

\bibitem[Gon{\c{c}}alves et~al., 2014]{gonccalves2014}
Gon{\c{c}}alves, L., Subtil, A., Oliveira, M.~R., and Bermudez, P. (2014).
\newblock {ROC} curve estimation: An overview.
\newblock {\em REVSTAT--Statistical Journal}, 12(1):1--20.

\bibitem[Gonz{\'a}lez-Manteiga et~al., 2011]{Gonzalez-Manteiga2011a}
Gonz{\'a}lez-Manteiga, W., Pardo-Fern{\'a}ndez, J.~C., and Van~Keilegom, I.
  (2011).
\newblock {ROC} curves in non-parametric location-scale regression models.
\newblock {\em Scandinavian Journal of Statistics}, 38(1):169--184.

\bibitem[Hsieh and Turnbull, 1996]{Hsieh1996}
Hsieh, F. and Turnbull, B.~W. (1996).
\newblock Nonparametric and semiparametric estimation of the receiver operating
  characteristic curve.
\newblock {\em Annals of Statistics}, 24(1):25--40.

\bibitem[In{\'a}cio and Rodr{\'i}guez-{\'A}lvarez, 2022]{InacioRdzAlvz2022}
In{\'a}cio, V. and Rodr{\'i}guez-{\'A}lvarez, M.~X. (2022).
\newblock The covariate-adjusted {ROC} curve: The concept and its importance,
  review of inferential methods, and a new {B}ayesian estimator.
\newblock {\em Statistical Science}, 37(4):541 -- 561.

\bibitem[Janes et~al., 2009]{Janes2009}
Janes, H., Longton, G., and Pepe, M.~S. (2009).
\newblock Accommodating covariates in receiver operating characteristic
  analysis.
\newblock {\em Stata Journal}, 9:17--39.

\bibitem[Janes and Pepe, 2009]{JanesPepe2009}
Janes, H. and Pepe, M.~S. (2009).
\newblock Adjusting for covariate effects on classification accuracy using the
  covariate-adjusted receiver operating characteristic curve.
\newblock {\em Biometrika}, 96(2):371--382.

\bibitem[Krzanowski and Hand, 2009]{Krzanowski2009}
Krzanowski, W.~J. and Hand, D.~J. (2009).
\newblock {\em {ROC} Curves for Continuous Data}.
\newblock Chapman \& Hall/CRC, Boca Ratón.

\bibitem[Mart{\'\i}nez-Camblor et~al., 2011]{Martinez-Camblor2011b}
Mart{\'\i}nez-Camblor, P., Carleos, C., and Corral, N. (2011).
\newblock Powerful nonparametric statistics to compare {$k$} independent {ROC}
  curves.
\newblock {\em Journal of Applied Statistics}, 38(7):1317--1332.

\bibitem[Mart{\'\i}nez-Camblor et~al., 2013]{Martinez-Camblor2013a}
Mart{\'\i}nez-Camblor, P., Carleos, C., and Corral, N. (2013).
\newblock General nonparametric {ROC} curve comparison.
\newblock {\em Journal of the Korean Statistical Society}, 42(1):71--81.

\bibitem[Mart{\'\i}nez-Camblor and Corral, 2012]{Martinez-Camblor2012}
Mart{\'\i}nez-Camblor, P. and Corral, N. (2012).
\newblock A general bootstrap algorithm for hypothesis testing.
\newblock {\em Journal of Statistical Planning and Inference}, 142(2):589--600.

\bibitem[Nakas et~al., 2023]{Nakas2023}
Nakas, C., Bantis, L., and Gatsonis, C. (2023).
\newblock {\em ROC Analysis for Classification and Prediction in Practice}.

\bibitem[Pardo-Fern{\'a}ndez et~al., 2014]{Pardo-Fernandez2014}
Pardo-Fern{\'a}ndez, J.~C., Rodr{\'i}guez-{\'A}lvarez, M.~X., and Van~Keilegom,
  I. (2014).
\newblock A review on {ROC} curves in the presence of covariates.
\newblock {\em {REVSTAT}--Statistical Journal}, 12(1):21--41.

\bibitem[Pepe, 2003]{Pepe2003}
Pepe, M.~S. (2003).
\newblock {\em The Statistical Evaluation of Medical Tests for Classification
  and Prediction}.
\newblock Oxford University Press, Oxford.

\bibitem[{R Core Team}, 2022]{ManR}
{R Core Team} (2022).
\newblock {\em R: A Language and Environment for Statistical Computing}.
\newblock R Foundation for Statistical Computing, Vienna, Austria.

\bibitem[Rodr{\'\i}guez-{\'A}lvarez and Roca-Pardi{\~n}as,
  2023]{ManRodriguez-Alvarez2017}
Rodr{\'\i}guez-{\'A}lvarez, M.~X. and Roca-Pardi{\~n}as, J. (2023).
\newblock {\em np{ROCR}egression: Kernel-Based Nonparametric {ROC} Regression
  Modelling}.
\newblock R package version 1.0-7.

\bibitem[Rodr{\'\i}guez-{\'A}lvarez et~al., 2011]{Rodriguez-Alvarez2011}
Rodr{\'\i}guez-{\'A}lvarez, M.~X., Roca-Pardi{\~n}as, J., and
  Cadarso-Su{\'a}rez, C. (2011).
\newblock {ROC} curve and covariates: extending induced methodology to the
  non-parametric framework.
\newblock {\em Statistics and Computing}, 21(4):483--499.

\bibitem[Rodr{\'\i}guez-{\'A}lvarez et~al., 2018]{Rodriguez-Alvarez2018}
Rodr{\'\i}guez-{\'A}lvarez, M.~X., Roca-Pardi{\~n}as, J., Cadarso-Su{\'a}rez,
  C., and Tahoces, P.~G. (2018).
\newblock Bootstrap-based procedures for inference in nonparametric
  receiver-operating characteristic curve regression analysis.
\newblock {\em Statistical Methods in Medical Research}, 27(3):740--764.

\end{thebibliography}

\end{document}